\documentstyle[preprint,aps,eqsecnum,fixes,epsfig,amstex]{revtex}

\makeatletter           
\@floatstrue
\def\figure{\let\@capwidth\columnwidth\@float{figure}}
\let\endfigure\end@float
\@namedef{figure*}{\let\@capwidth\textwidth\@dblfloat{figure}}
\@namedef{endfigure*}{\end@dblfloat}
\makeatother

\begin {document}

\tightenlines
\newcommand{\nc}{\newcommand}
\nc{\bea}{\begin{eqnarray}}
\nc{\eea}{\end{eqnarray}}
\nc{\beqa}{\begin{eqnarray}}
\nc{\eeqa}{\end{eqnarray}}
\nc{\NCS}{N_{\rm CS}}
\nc{\lsim}{\mbox{\raisebox{-.6ex}{~$\stackrel{<}{\sim}$~}}}
\nc{\gsim}{\mbox{\raisebox{-.6ex}{~$\stackrel{>}{\sim}$~}}}
\nc{\mD}{m_{\rm D}}
\nc{\meq}{m_{\rm eq}}
\nc{\mth}{m_{\rm th}}


\preprint {UW/PT 00-01}

\title{ Sphaleron rate in the symmetric electroweak phase} 

\author {Guy D. Moore}

\address
    {%
Department of Physics,
University of Washington, 
Seattle WA 98195-1560 USA	
    }%
\date {January 2000}

\maketitle
\vskip -20pt

\begin {abstract}%
{%
Recently B\"{o}deker has presented an effective infrared theory for the
dynamics of Yang-Mills theory, suitable for studying the rate of baryon
number violation in the early universe.  We extend his theory to include
Higgs fields, and study how much the Higgs affects the baryon number violation 
rate in the symmetric phase, at the phase coexistence temperature of a
first order electroweak phase transition.  The rate is about 20\%
smaller than in pure Yang-Mills theory.
We also analyze the sphaleron rate in the analytic crossover regime.
Our treatment relies on the ergodicity conjecture for 3-D scalar
$\phi^4$ theory.
}%
\end {abstract}

\thispagestyle{empty}


\section {Introduction}

Baryon number is violated in the standard model \cite{tHooft}, and while 
the rate of the violation is negligibly small in vacuum, it can be
significant at higher temperatures \cite{Kuzmin,ArnoldMcLerran}.

Recently there has been substantial interest in how fast baryon number
is violated in the standard model at high temperatures.  This is both
because the rate is important for baryogenesis, and because it is an
interesting problem in the more general framework of understanding
dynamics of hot gauge theories.

The last 3 years have seen major progress in this problem.  First,
analytic work has clarified what physics is relevant.  While it has long 
been known that infrared or ``soft'' 
gauge fields with momenta $p \sim g^2 T$ are
responsible for baryon number violation \cite{ArnoldMcLerran}, the
recent work of Arnold, Son, and Yaffe has shown that ``hard'' modes with 
momenta $p \sim T$ play an essential role in modifying the dynamics of
the $p \sim g^2T$ degrees of freedom \cite{ASY,HuetSon,Son}.
B\"{o}deker has gone further, showing the role of scatterings between
such hard modes by exchange of $g^2T \lsim p \lsim gT$ modes
\cite{Bodeker,moreBodeker,moreASY}.  There has also been progress in
numerically studying the effective infrared (IR) theories to determine
the size of baryon number violation \cite{particles,Wfields}.

The rate of baryon number violation is characterized by the sphaleron
rate\footnote{Our normalization is $D_i = \partial_i + iT^a A^a_i$, and
$F_{ij} = [D_i , D_j]$.  $\alpha$ means $\alpha_w$, 
and $g$ means $g_w$.}
\bea
\Gamma & \equiv & \lim_{V,t \rightarrow \infty} \frac{\langle(\NCS(t)
	-\NCS(0))^2\rangle}{Vt} \, , \\
\NCS(t)-\NCS(0) & = & \frac{1}{16 \pi^2} \int_0^t \int d^3x 
	\epsilon_{ijk} E_i^a F_{jk}^a \, .
\eea
B\"{o}deker has shown that in pure Yang-Mills theory, $\Gamma$ has the
parametric form \cite{Bodeker}
\begin{equation}
\Gamma = \kappa ' \left( \log \frac{\mD}{g^2T} + C + O(1/\log) \right)
	\left( \frac{g^2T^2}{\mD^2} \right) \alpha^5 T^4 \, .
\label{parametric_form}
\end{equation}
The constant\footnote{Actually it is not a constant, but contains a
$\log(\log(m_D/g^2T))$, see Eq. (\protect{\ref{value_of_sigma}}).} 
$C$ can and recently has been determined
analytically \cite{AY2_long}.  The leading coefficient $\kappa'$ can be
obtained numerically by studying a local, UV finite theory, which is
precisely Langevin dynamics for classical 3 dimensional nonabelian gauge 
fields:
\bea
D_\tau A_i^a(x,\tau) & = & - g^2 \frac{\partial H_A(A(\tau))}{\partial
	A_i^a(x,\tau)} + \xi_i^a(x,\tau) \, , \\
H_A & = & \int d^3x \frac{1}{4g^2} F_{ij}^a F_{ij}^a(x) \, , 
	\quad -g^2 \frac{\partial H_A}{\partial A_i^a} = 
	(D_j F_{ji})^a = (D\times B)_i^a \, , \\
\langle \xi_i^a(x_1,\tau_1) \xi_j^b(x_2,\tau_2)\rangle & = & 
	2 g^2T \delta(x_1-x_2) \delta(\tau_1-\tau_2) \delta_{ij}
	\delta^{ab} \, ,
\eea
Here $D_\tau$ is the covariant derivative in Langevin time $\tau$, $F_{ij}^a$
the (nonabelian) magnetic field, and $\xi$ Gaussian white noise
normalized as shown.\footnote{Note that the Langevin time $\tau$ has
dimensions of length squared, and recall our scaling convention for
$A$.}.  

The form of this theory is identical to the nonabelian Ampere's law, but 
with the current replaced with ${\vec j} = \sigma \vec{E}$, as we would
expect in a conducting medium;
\begin{equation}
D \times B = - \sigma E \; \;{\rm (\: + \; noise)}\, ,
\end{equation}
with the Langevin time identified as $\tau = \sigma t$.  Here $\sigma$
is a ``color conductivity'' which describes the current response to very 
infrared external fields.  It can be treated as a constant, to (next to) 
leading order in $\log(1/g)$, because the mean free path for color
changing collisions is shorter than the $1/g^2 T$ scale where $\Gamma$
is set, by a factor of $\log(1/g)$.  An explicit expression for $\sigma$ 
at next to leading log order, found by Arnold and Yaffe,
\cite{AY2_short}, will be
presented in Eq. (\ref{value_of_sigma}).
$\kappa'$ is determined in this theory as
\begin{equation}
\kappa' = \frac{3\sigma}{2 \pi} \lim_{V,\tau \rightarrow \infty}
	\frac{\langle( (\NCS(\tau) - \NCS(0))^2 \rangle}{V\tau} \, .
\end{equation}
This effective theory was studied numerically in \cite{bodek_paper},
with the result (see below) that $\kappa' \simeq 10$.
This effective theory is not sufficient to determine the $O(1/\log)$
coefficient in Eq. (\ref{parametric_form}), 
and neither will anything we discuss.

Most of what we have described, and in particular everything in
references 
\cite{ASY,Bodeker,moreBodeker,moreASY,particles,Wfields,AY2_long,bodek_paper,AY2_first},
is for the case
of pure Yang-Mills theory.  This is appropriate if we are interested in
temperatures very much higher than the electroweak phase transition (or
crossover) temperature, because in that case the Higgs boson has a large
thermal mass and can be removed perturbatively.  However, the main
application we are interested in, baryogenesis, requires knowing
$\Gamma$ in a theory with at least one Higgs boson, in the symmetric
phase but at or slightly
below the equilibrium temperature $T_{\rm eq}$ 
for the electroweak phase transition.  It is not clear whether
discarding the Higgs physics is justified in this case, and we should
rethink both the appropriate effective theory, and the numerical
determination of $\Gamma$, in this light.

In this paper we will attempt to fill this gap.  First we discuss how
important we expect Higgs physics to be by considering the thermodynamics 
of Yang-Mills Higgs fields in Section \ref{thermo}.  Then we construct
an appropriate infrared (IR) effective theory which generalizes
B\"{o}deker's effective theory in Section \ref{ef_theory}.  We discuss
the numerical implementation in Section \ref{numerics} and present
numerical results in \ref{result_sec}.  We find that, where the phase
transition is strong enough to preserve baryon number after its
completion, the change in $\Gamma$ due to Higgs physics is quite a small 
effect, so the error in using pure Yang-Mills theory is small, of order
20\%.  While we work in the minimal standard model (at experimentally
excluded values of the Higgs mass, to get a strong enough phase
transition), we expect the results to hold as well in extensions such as 
the MSSM with a light scalar top quark.  This conjecture could be tested 
by simulations in that theory, along the lines of what we do here.

\section{Thermodynamic influence of the Higgs}
\label{thermo}

To get an idea of how important Higgs physics will be for the sphaleron
rate, we will try to get an idea of how important it is
thermodynamically for the infrared transverse gauge boson
excitations which we expect to be responsible for baryon number
violation.  As we will see, this in fact gives a reasonable estimate for 
what difference the Higgs will make in the sphaleron rate.

To a very good approximation the thermodynamics of infrared bosonic
fields in the hot electroweak theory can be described by a three
dimensional path integral \cite{KLRS}.  In fact this can be
understood as a special case of the statement that the IR physics is
essentially classical, since the three dimensional path integral we
arrive at coincides with the partition function of the classical bosonic 
theory.  Up to parametrically suppressed corrections\footnote{The 3-D
theory should be viewed as an IR effective theory for the
thermodynamics below the scale $T$.  
At some level of precision it becomes necessary to
include high dimension operators.  If we are interested in
thermodynamics at the length scale $1/g^2T$, then neglecting the high
dimension operators causes errors of $O((g^2T/T)^2)$ times an additional 
explicit factor of $\alpha$ because the high dimension operators are
radiatively induced, leading to an $\alpha^3$ error.  Of course, on less
infrared scales the effective theory is less accurate.},
the partition function governing the thermodynamics is
\cite{KLRS}
\bea
Z & = & \int {\cal D}A_i {\cal D}A_0 {\cal D} \Phi \exp -H/T \, , \\
H & = & \int d^3x \frac{1}{4g^2} F_{ij}^a F_{ij}^a + 
	\frac{1}{2g^2} (D_i A_0)^a (D_iA_0)^a 
	+ \frac{\mD^2(T)}{2g^2} A_0^a A_0^a
	+ \frac{\lambda_A}{4 g^4} (A_0^a A_0^a)^2 \nonumber \\ &&
	\qquad + (D_i \Phi)^\dagger (D_i\Phi) + 
	m_H^2(T) \Phi^{\dagger} \Phi + \lambda (\Phi^\dagger \Phi)^2
	+ \frac{\lambda_{A\Phi}}{2g^2} A_0^a A_0^a \Phi^\dagger \Phi \, .
\label{H_DR}
\eea
The couplings $\lambda$, $g^2$, $\lambda_A \sim \alpha^2$, 
and $\lambda_{A\Phi} =
g^2/2 + O(\alpha^2)$ are determined by a matching calculation; $g^2$,
$\lambda$, and the field wave function normalizations
correspond to those of the 4-D theory at a renormalization point
given roughly by $T$.  We will always treat $\lambda \sim g^2$ for power 
counting purposes, as is appropriate given the renormalization structure 
of the theory.  It is sometimes useful to consider $\lambda \ll g^2$ and 
to expand in $\lambda/g^2$, but if we do so it is implied that $\lambda$ 
is still $\gg g^4$, for instance.
$A_0$ is the remnant of the temporal connection and can be thought 
of as an adjoint scalar field.  Its mass term is the Debye mass
responsible for charge screening, and is given by
\begin{equation}
\mD^2(T) = \frac{g^2 T^2}{12} ( 4N_{\rm c} + N_f + 2 N_s ) 
	- {\rm \, 3D \; counterterm \,} + O(g^4T^2) \, ,
\end{equation}
with $N_{\rm c}=2$ the second Casimir of the group, $N_f$ the number of
chiral, fundamental representation fermions, and $N_s$ the number of
fundamental representation complex scalars.  In the minimal standard
model, the $O(g^2)$ piece is $\mD^2 = (11/6) g^2 T^2$.

Because of the Debye mass, the $A_0$ field is heavy and has little
influence on the very infrared thermodynamics, leading only to a
rescaling of the effective gauge coupling by an $O(g)$ correction
\cite{KLRS}.  However the
same is not generically true for the Higgs field.  This is because,
unlike the $A_0$ field, it has a negative vacuum mass squared which may
approximately cancel the positive induced thermal mass;
\begin{eqnarray}
&& m_H^2(T) = \mth^2 + m_{\rm vac}^2 \, , \qquad
	m_{\rm vac}^2 < 0 \, , \quad \mth^2 \sim g^2T^2 \, , \\
&& \left({\rm specifically, \;} \mth^2= \frac{(3 g^2 + {g'}^2 + 4
y_t^2+8\lambda )T^2}{16} \; {\rm in \; the \; MSM} \right)
\, . \nonumber 
\end{eqnarray}
At very large temperatures, $m_H^2(T) \sim g^2 T^2$ and its
thermodynamic effects are parametrically suppressed.  However, the
electroweak phase transition occurs where the positive thermal mass
squared and the negative tree one cancel up to $O(g^4T^2)$ corrections,
and it is in this regime that we need to know the sphaleron rate.  It
makes sense, then, to treat $m_H^2(T)$ parametrically as $O(g^4T^2)$,
though depending on the strength of the phase transition, it may either
be ``large'' or ``small'' within this parametric range.

\begin{figure}[tbp]
\centerline{\epsfxsize=4in\epsfbox{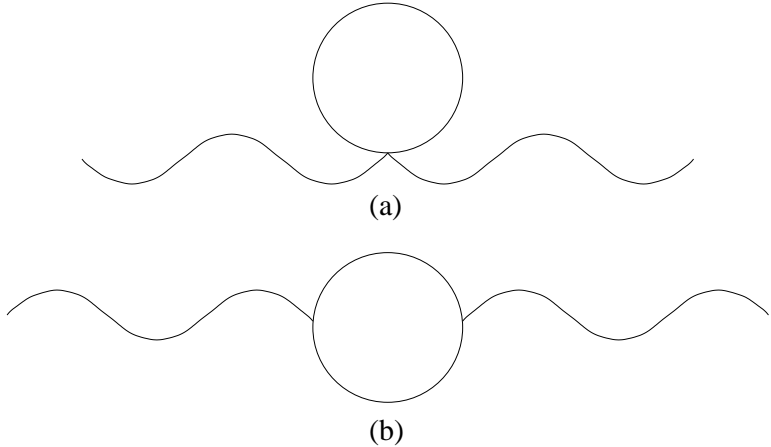}}
\caption{gauge field self-energy insertions generated by the Higgs
fields. \label{diagrams}}
\end{figure}

Let us now consider the thermodynamic influence of the Higgs boson on
the Yang-Mills fields.  Since we are only making estimates here, we will 
be content with one loop calculations even where the loop expansion is
not very reliable.  We can estimate the Higgs field's importance by
considering Higgs contributions to the gauge field self-energy at soft
external momentum $p \sim g^2T$.  The sum of diagram (a) in
Fig. \ref{diagrams} and any $p$ independent contributions from (b)
vanish in any gauge invariance respecting regularization; the
scalar contribution to the gauge field self-energy is then 
\begin{equation}
\Pi_{ij}^{ab}(p) = \frac{g^2T \delta^{ab}}{2} \int 
	\frac{d^3k}{(2\pi)^3} \left( 
	\frac{2 k_i 2 k_j}{( (k-\frac{p}{2})^2 + m_H^2) 
	( (k+\frac{p}{2})^2 + m_H^2)}
	- \frac{2k_i2k_j}{(k^2+m_H^2)^2} \right) \, .
\label{Higgs_corr}
\end{equation}
If we take $p \ll m_H$, the result is
\begin{equation}
\Pi_{ij}^{ab}(p) = - \frac{g^2T \delta^{ab}}{48 \pi m_H} 
	\left( \delta_{ij} p^2 - p_i p_j \right) \, ,
\label{heavy_Higgs}
\end{equation}
whereas in the regime where $m_H \ll p$, we get
\begin{equation}
\Pi_{ij}^{ab}(p) = - \frac{g^2 T\delta^{ab}}{32|p|} 
	\left( \delta_{ij} p^2 - p_i p_j \right) \, .
\label{light_Higgs}
\end{equation}
In either case the sign of the effect is such that it reduces the size
of infrared gauge field excitations; so the presence of a Higgs field
should reduce $\Gamma$.

The calculation is for the symmetric phase, where there is no Higgs
condensate.  When there is a condensate $\phi_0$ (normalized as
$\phi_0^2 = \langle 2 \Phi^\dagger \Phi \rangle$) then there is an
induced mass for the gauge bosons of $m_W^2 = g^2 \phi_0^2/4$, which
will of course further reduce $\Gamma$.  However, we are primarily
interested right now in the symmetric phase case.

To decide how important the scalar contribution is, we need to know 
what momentum $p$ is characteristic for the baryon number violating
processes we intend to study.  Since the physics we are after is
nonperturbative physics, we expect the answer to be, $p$ such that 
perturbation theory is breaking down.  We can estimate what $p$ is
necessary by computing the contribution of gauge bosons to the
self-energy.  The result is not gauge invariant, but in Landau gauge the 
analogous contribution from gauge and ghost loops is \footnote{In
Feynman gauge the $11$ becomes $14$.  The answer differs from that in
\protect{\cite{Kajantie_old}}, for instance, because we have left out
the (heavy) $A_0$ field, treated as light there.}
\begin{equation}
\Pi_{ij}^{ab}(p) {\rm \; from \; bosons \,} = \frac{11 N_{\rm c} g^2 T
	\delta^{ab}}{64|p|} \left( \delta_{ij} p^2 - p_i p_j \right) \, .
\end{equation}
The sign is the opposite, indicating that gauge self-interactions lead
to a more rapid onset of nonperturbatively large fields.  The loop
contribution comes on order the tree inverse propagator $p^2 \delta_{ij} 
- p_i p_j$, and the calculation therefore breaks down completely, for $p 
\sim g^2T/3$.  We take this as a fair estimate of the nonperturbative
scale, though such an estimate certainly cannot be considered accurate
to better than about a factor of 2.

For a heavy Higgs boson, then, we expect a correction of order 
$(g^2T/m_H(T))/(48 \pi)$; but even for $m_H(T) \simeq 0$ we only expect
a correction from the Higgs boson to the propagator at the relevant
momentum to be of order $1/11$.  This corresponds to a rescaling by
$\sim (1/11)$ of the 
length scale where perturbation theory breaks down.  However, since the
fifth power of this length scale enters the sphaleron rate, the
importance of the Higgs boson would be a little larger, as large as
a $(10/11)^5 \simeq 0.6$ reduction of $\Gamma$.  This analysis also
implies that, when there is a condensate, it starts to significantly
suppress baryon 
number violation when the induced gauge field mass is of order $g^2T/3$, 
which requires $\phi_0 \simeq 2 gT/3$.  Of course for such a small
condensate, the perturbative notion of ``condensate'' is lost and what
we are doing is unreliable; but this should give us some estimate of how 
strong the transition needs to be before the drop in the sphaleron rate
becomes significant.

In the standard model and in the regime where the phase transition is
strong, we can estimate the symmetric phase $m_H(T_{\rm eq})$; at
leading order in $\lambda/g^2$, a standard calculation from the
curvature of the one loop effective potential gives
\begin{equation}
m_H(T_{\rm eq}) = g^2T / (16 \pi \sqrt{\lambda/2g^2}) \, .
\end{equation}
Plugging this into Eq. (\ref{heavy_Higgs}) gives a self-energy
correction of $-\sqrt{\lambda / 18 g^2}$, and a correction to the
sphaleron rate of 
\begin{equation}
\frac{\delta \Gamma}{\Gamma} = - 5 \sqrt{\frac{ \lambda}{18 g^2}} \, .
\label{dgamma}
\end{equation}
Below we will study the case
where $\lambda / g^2 = 0.036$, for which the correction is $22\% $.
However, this mass is not in the regime $m_H \gg p$ for $p=g^2T/3$, so
we have computed Eq. (\ref{Higgs_corr}) numerically for this $m_H$ and
$p$; it yields $-0.037$.  A rough estimate is that we will see $\Gamma$
reduced to $(1-.037)^5 = .83$ of the Yang-Mills theory value.  
This estimate is quite rough, but it
gives the idea that the symmetric phase rate will be lower, but not much 
lower, than the pure Yang-Mills theory rate.
%
%

To conclude, at the thermodynamic level, one would expect the Higgs
field induced correction
to infrared gauge field behavior, and hence to the sphaleron rate,
to be rather small in the symmetric
phase, and  of order 1 for condensates smaller than $\phi_0 \simeq (2/3)
gT$.  Naturally, for larger condensates the suppression becomes very
substantial, see \cite{ArnoldMcLerran,broken_nonpert}.

\section{Effective Theory in the Presence of the Higgs}
\label{ef_theory}

Now we will go about constructing appropriate effective theories for
studying baryon number violation when there is a light Higgs boson.  
Our treatment will follow very closely that of
\cite{HuetSon,Bodeker,ASY}.  Indeed, most of the complications stem from 
the Yang-Mills sector and have been resolved in the literature cited;
the Higgs boson will introduce comparatively minor new complications.

We will also leave out the U(1) field in what follows, even though it is 
light.  This is a common and probably reasonable approximation.  We make
it partly because we expect the influence of the U(1) field to be weak;
there is no direct interaction between the SU(2) and U(1) gauge fields,
and the U(1) physics makes rather small modifications to the
thermodynamics of the SU(2)-Higgs system
\protect{\cite{KLRS_U1}}.  Also, including the U(1)
physics would prove to be a significant complication, since at the $k
\sim g^2T$ scale the U(1) fields are overdamped but with a $k$ dependent
damping which must be treated as a nonlocal effect even in the final
effective theory.  We will not try to address this problem here.

Since we are not concerned with cases where there is a very large
condensate $\phi_0 \sim T/g$, degrees of freedom with momentum $k \sim
T$ behave at leading order as massless free fields propagating 
in the background of the IR fields, and the degrees of
freedom with $k \leq gT$ have large occupation numbers and can be
treated as classical fields.  The infrared behavior of the theory is
described by a classical effective theory in which the $k \sim T$
degrees of freedom have been analytically integrated out.  The resulting 
equations of motion are similar to those of the classical field theory.
Defining
\beqa
\label{classical_H}
H & = & H_A + H_\Phi + H_E + H_\Pi \, , \\
\label{H_A}
H_A & = & \int d^3x \frac{1}{4g^2} F_{ij}^a F_{ij}^a(x) \, , \quad
	F_{ij} \equiv [D_i , D_j] \, , \\
\label{H_phi}
H_\Phi & = & \int d^3x (D_i \Phi)^\dagger (D_i \Phi)(x) + 
	m_H^2 \Phi^\dagger \Phi(x) + \lambda (\Phi^\dagger \Phi)^2(x) \, , \\
\label{H_E}
H_E & = & \int d^3x \frac{1}{2g^2} E_i^a E_i^a(x) \, , \quad
	E_i \equiv [D_t , D_i] \, , \\
\label{H_Pi}
H_\Pi & = & \int d^3x \frac{1}{2} \Pi^\dagger \Pi(x) \, , \quad
	\Pi \equiv D_t \Phi \, ,
\eeqa
the effective field equations will be the classical equations of
motion derived from this Hamiltonian, supplemented by the
hard thermal loop effects which arise from the integration over
the heavy modes.  The hard thermal loop (HTL) field equations are
\cite{BraatenPisarski,HTLpapers,old_Iancu,old_Nair}
\bea
\label{gauge_gT}
(D_t E_i)^a(x) & = & - g^2 \frac{\partial H_A}{\partial A_i^a(x)} 
	- g^2 \frac{\partial H_\Phi}{\partial A_i^a(x)}
	- \mD \int \frac{d\Omega_v}{4\pi} v_iW^a(x,v) \, , \\
\label{Higgs_gT}
D_t \Pi(x) & = & - \frac{\partial H_\Phi}{\partial \Phi^\dagger(x)} 
	- \mth^2 \Phi(x) \nonumber \\
& = & D_i D_i \Phi(x) - \left(m_H^2(T)
	-2 \lambda \Phi^\dagger \Phi(x)\right) \Phi(x)	\, , \\
\label{W_gT}
(D_t W)^a(x,v) & = & - v_i (D_i W)^a(x,v) + \mD v_i E^a_i(x) \, , \\
(D_i E_i)^a (x) & = & g^2 \left( \Pi^\dagger iT^a \Phi(x) 
	+ {\rm c.c.} \right)
	+ \mD \int \frac{d\Omega_v}{4\pi} W^a(x,v) \, .
\label{Gauss_gT}
\eea
These equations determine the field evolution, given initial information 
for $A$, $E$, $\Phi$, $\Pi$, and $W$, up to the 
freedom to choose the time dependent
gauge.  The last equation is Gauss' Law and needs to be applied as a
constraint on the initial conditions; it commutes with the other
equations so it remains valid at later times.

The only hard thermal loop effect on the soft scalar field is a thermal
mass squared correction.  We discuss this point at some length in
Appendix \ref{App_HTL}, where we demonstrate the absence of all other HTL
effects involving scalar external lines.  We
write the correction as $\mth^2$ (the thermally induced mass
squared) and the sum $m_H^2 + \mth^2 = m_H^2(T)$.  In what follows 
we will absorb this shift in $m_H^2$ into $H_{\Phi}$.
This does not require any modification of Eq. (\ref{gauge_gT}) because
$\mth^2 \Phi^\dagger \Phi$ is independent of $A$.

The HTLs for gauge fields are 
much more complicated, and are implemented here with auxiliary fields,
the $W$ fields, which allow them to be written in a local way
\cite{old_Iancu,old_Nair}.  The penalty is that the $W$ fields depend on 
direction $v$ as well as position $x$.  Here $v$ is a unit vector, and
$d\Omega_v$ is an integral over directions (the unit sphere) 
normalized so $\int d\Omega_v / 4 \pi = 1$.
We normalize $W$ slightly differently than the references, absorbing a
factor of $m_D$ into its normalization so it 
enters the Hamiltonian with the same weight as the gauge fields.  Its
listing in Eq. (\ref{classical_H}) would be
\begin{equation}
H_W = \int d^3x \frac{1}{2g^2} \int \frac{d\Omega_v}{4\pi} W^a(x,v)
	W^a(x,v) \, .
\end{equation}
It is possible to derive the field equations from this (generalized) 
Hamiltonian,
together with Eq. (\ref{classical_H}), but it requires rather nontrivial 
Lie-Poisson brackets \cite{Iancu_new}.  Note that the hard scalar
contributions to the gauge HTL is identical in structure to that from
hard gauge and fermionic degrees of freedom.  This is most easily seen
in the kinetic theory derivation of the gauge HTL's \cite{old_Iancu}, where
one finds that the spin of a hard degree of freedom only matters at
subleading order in $g$.

Eqs. (\ref{gauge_gT})--(\ref{Gauss_gT}) already give an effective theory
which is amenable to numerical treatment along the lines of that given
for pure Yang-Mills theory in \cite{particles,Wfields}.  The added
complication of including the Higgs field is much less than that
involved in treating the hard thermal loops.  However, such a numerical
implementation is not ideal, because the system of equations presented
is not UV finite.  In particular the classical gauge and scalar degrees
of freedom, with dynamics determined by Eqs. (\ref{gauge_gT}) and
(\ref{Higgs_gT}), generate UV divergent loop corrections which can be
considered as extra contributions to the hard thermal loops.  In a
lattice regularization, these extra contributions are finite but grow
linearly as the lattice spacing is made smaller, and are not rotationally 
invariant \cite{Smilga}.  So long as $m_D$ is kept suitably large 
this should be a subdominant effect, but it is not clear that it is safe 
to make $m_D$ as large as the inverse lattice spacing.  Because of these 
problems it is not easy to extract high precision information from such
a simulation, and if we expect a fairly small change to $\Gamma$ due to
the Higgs bosons, it may be problematic to isolate it from lattice
spacing effects.  Rather, we will follow B\"{o}deker \cite{Bodeker} and
integrate out more degrees of freedom to construct what ultimately
proves to be a simpler and cleaner effective theory (though no longer
valid to corrections parametrically suppressed by a full power of $g$),
which will prove a better test-bed for studying the importance of Higgs
physics.  

With this in mind we consider integrating over the $gT$ scale, down to
some intermediate scale $\mu \ll gT$.  We are also removing physics with 
a frequency scale $\omega \sim gT$; however we cannot directly consider
only degrees of freedom with $\omega \ll k \ll gT$; for now $\omega$ is
permitted to be as large as $\mu$.

Here we will only treat the case in which there is not a Higgs boson
condensate of order $\phi_0 \sim \mu / g$.  This either restricts how
deeply into the broken phase our analysis remains valid, or how small we 
are permitted to make the scale $\mu$ which we integrate down to.  We do 
not consider this restriction problematic because we are mostly
interested in the symmetric phase, and when there is a large condensate
we will need to use other tools to determine the rate anyway.  We have
already considered the broken phase problem in \cite{broken_nonpert}.

Under this assumption, the behavior of gauge bosons with $k \gg \mu$ is
not significantly changed by a Higgs condensate.  It is also not
significantly changed by the background of IR Higgs fluctuations, as can 
be verified by a loopwise analysis like the one in the last section; and 
the effects of hard Higgs excitations are already included by the HTL
effective action.  Therefore the integration over these degrees of
freedom proceeds as it does in the case without Higgs fields.
The integration over gauge and $W$ field degrees of freedom is very
nontrivial and has been treated at length by B\"{o}deker
\cite{Bodeker,moreBodeker} and by Arnold, Son, and Yaffe
\cite{moreASY,AY2_long}.  They show that a collision 
term is induced for the $W$ fields, together with noise required by the
fluctuation dissipation theorem.  The integration over the $gT$ scale
Higgs fields is much simpler.  The Higgs field equation of
motion, Eq. (\ref{Higgs_gT}), is the same below the $gT$ scale as it is 
up to the $T$ scale.  Provided that we keep $\mu \gg m_H(T)$, the Higgs
field still undergoes free relativistic propagation in the soft gauge
and Higgs field background up to corrections $O(g^2T/\mu)$.  Such
propagation is precisely what generated the hard thermal loops.
Standard power counting shows that contributions from the Higgs fields
with momentum $k> \mu$ to more IR degrees of freedom 
are suppressed by $g^2T/\mu$ except for UV divergent contributions.
But, as shown in \cite{Nauta}, the structure of UV divergences in the
classical field theory coincides exactly with the hard thermal loops.
Hence the $k > \mu$ Higgs degrees of freedom induce HTL's and
parametrically suppressed additional effects.  
The HTL effects from the $gT$ Higgs
fields are smaller by a factor of $g$ than those from $T$ physics, so we
can neglect them next to the HTL effects already being included.

It is worth remarking why physically no collision integral is induced
when we integrate out scalars with momentum $k \sim gT$.  The total
scattering rate for hard modes by exchange of a soft gauge particle is
$\sim g^2 T \log(1/g)$, with the logarithm arising from the momentum
region $gT$ to $g^2T$.  This large collision rate originates from the IR 
singular $s^2/t^2$ matrix elements in gauge boson exchange.
When we integrate out $gT$ bosons we must
explicitly include their contribution to scatterings via a collision
integral.  On the other hand, no scattering process between hard modes
which is mediated by single scalar exchange has a scattering rate
greater than $\sim g^4 T \log(T/m_H)$.\footnote{This can be 
verified by looking
at the matrix elements of tree level $2 \rightarrow 2$ scattering
processes mediated by a scalar, none of which contain $s^2/t^2$ type
terms.  At worst they go as $s^2/tu$ and are log IR divergent.}  
Such rare scatterings are not
important at leading order because we are eventually 
interested in physics at the $1/g^2T$ length scale.

On length scales more infrared than $1/gT$, Eqs. (\ref{gauge_gT}) and
(\ref{W_gT}) describe overdamped evolution for the gauge fields.
Further, the $W$ field equation of motion is linear in its slowly
varying source $\propto v_i E_i$.  Therefore it is 
permissible at leading order in $\mu / gT$ 
to drop both the $D_t E$ term in Eq. (\ref{gauge_gT})
and the $D_t W$ term in Eq. (\ref{W_gT}) \cite{HuetSon,Son,Bodeker},
leading to an effective theory for the $\mu \ll gT$ scale physics,
\footnote{For a discussion of the last equation, see
\protect{\cite{AY2_first}}.}
\bea
\label{gauge_mu}
\mD \int \frac{d\Omega}{4\pi} v_iW^a(x,v) & = & 
	-g^2 \frac{\partial H_A}{\partial A_i^a(x)}
	- g^2 \frac{\partial H_\Phi}{\partial A_i^a(x)} \, , \\
\label{Higgs_mu}
D_t \Pi(x) & = & -\frac{\partial H_\Phi}
	{\partial \Phi^\dagger(x)} \, , \\ 
\label{W_mu}
v_i (D_i W)^a(x,v) & = & \mD v_i E_i^a(x) 
	- \int \frac{d\Omega_{v'}}{4\pi} C_{vv'} W^a(x,v') 
	+ \zeta^a(x,v) \, , \\
m_D \int \frac{d\Omega_v}{4\pi} W^a(x,v) & = & 0 \, .
\eea
The new features are the collision integral $C_{vv'}$, whose form and
value is discussed in \cite{Bodeker,moreASY,AY2_long}, and the noise
$\zeta$.  At leading order in $\log(gT/\mu)$ the collision integral is 
\begin{equation}
C_{vv'} \simeq \gamma \left( \delta_{S^2}(v-v') - \frac{4}{\pi} 
	\frac{(v\cdot v')^2}{\sqrt{1 - (v \cdot v')^2}} \right) \, ,
	\qquad \gamma \simeq \frac{N_c g^2T}{4\pi} 
	\log \frac{m_D}{\mu} \, .
\end{equation}
The noise is Gaussian and white with two-point correlator
\begin{equation}
\langle \zeta^a(x,v,t) \zeta^b(y,v',t')\rangle = 2g^2T C_{vv'} 
	\delta^{ab} \delta(x-y) \delta(t-t') \, .
\end{equation}
Because of our nonstandard $W$ field normalization, the normalization of 
the noise differs from the references.  Note that the size of the collision
integral is parametrically $C_{vv'} \sim g^2T \log(gT/\mu)$.

At the $k \sim g^2T$ scale this effective theory has two natural time
scales.  First there is the time scale $\sim 1/g^2T$, on which
Eq. (\ref{Higgs_mu}) allows the Higgs fields to evolve.  There is also a 
scale, $\sim 1/g^4T \log(1/g)$, 
on which the overdamped gauge and $W$ fields evolve.  This reflects the
fact that the gauge HTL's include Landau damping and lead to
overdamped evolution, while the Higgs HTL's are just a mass correction 
and do not induce any $g^2T^2$ size damping, so the Higgs field
is not overdamped.

If we view the system on length scales $\sim 1/g^2T$ and on the Higgs
time scale $\sim 1/g^2T$, the gauge fields look ``frozen'' up to
parametrically suppressed effects.  On these time scales the Higgs field 
evolution is determined by Eq. (\ref{Higgs_mu}) with the gauge field
background frozen.  What is the behavior of such a system?  The Higgs
propagates on an inhomogeneous background connection, interacting with
itself via the nonlinear quartic coupling term.  There is a widely
(though not universally) believed conjecture that the evolution of a
scalar field theory with quartic self-coupling in 3 dimensions should be 
ergodic, in which case it will randomize itself on the $1/g^2T$
time scale (since there is no other available time scale for its
evolution).  This view is supported, for
instance, by the calculation of the damping rate of a scalar at rest, in 
a flat connection but with a quartic interaction; the damping rate is
parametrically  $\sim \lambda^2 T^2 / m_H \sim g^2T$, and the damping
arises primarily from degrees of freedom with $k \sim m_H$
\cite{Aarts}.  We expect that the inhomogeneous connection should only
make the randomization of the scalar field more efficient.  In
particular we speculate that the spectrum of the $D^2$ operator for a
typical 3-D gauge field background exhibits Anderson localization at all 
frequencies.

On the $1/g^4T$ time scale on which the gauge field evolves, the Higgs
field will thoroughly explore its fixed connection thermal ensemble.
Therefore, on the time scale on which $A$ evolves, it
will see a thermodynamic average of the possible Higgs field
configurations.  We emphasize that we are relying here on the
conjectured ergodicity of 3-D scalar $\phi^4$ theory.  It is also not
clear that our treatment 
will remain true very near the endpoint of the electroweak 
phase transition, where the Higgs field correlation length grows to be
$\gg 1/g^2T$, because in this regime it may take much longer for the
Higgs field to explore its fixed gauge field ensemble.
We will exclude that regime from consideration, although
it is not clear to us that the effective theory we will derive
cannot be used there as well.

When our assumption is valid, we should average the $\Phi$ dependent
part of Eq. (\ref{gauge_mu}) over the Higgs thermal ensemble,
\begin{equation}
-g^2 \frac{\partial H_\Phi}{\partial A_i^a} \rightarrow
-g^2 \left\langle \frac{\partial H_\Phi}{\partial A_i^a} \right\rangle 
= g^2T \frac{\partial}{\partial A_i^a} \log \int {\cal D\Phi}
\exp(-H_\Phi/T) \, .
\end{equation}
The RHS of Eq. (\ref{gauge_mu}) can be understood as $g^2T$ times a
variation of a 3-D effective action describing the thermodynamics of 
the gauge fields.  The sole
modification from the inclusion of Higgs fields is that the effective
action should include, besides the gauge part, a nonlocal piece
arising from integrating over the Higgs fields.  Hence the effective
theory describing infrared gauge bosons is
\bea
\label{gauge_noH}
\mD \int \frac{d\Omega}{4\pi} v_iW^a(x,v) & = & - g^2T 
	\frac{\partial}{\partial A_i^a(x)} 
	\frac{H_{\rm eff}(A)}{T} \, , \\
\label{def_Heff}
\frac{H_{\rm eff}(A)}{T} & = & - \log \int {\cal D}\Phi \exp 
	\left( - \frac{H_A + H_{\Phi}}{T} \right) \, , \\
\label{W_noH}
v_i (D_i W)^a(x,v) & = & \mD v_i E_i^a(E) 
	- \int \frac{d\Omega_{v'}}{4\pi} C_{vv'} W^a(x,v') 
	+ \zeta^a(x,v) \, .
\eea
This is identical with the ``theory 2'' of reference
\cite{AY2_short,AY2_long}, except
for the Higgs field additions to $H_{\rm eff}$.

This theory is probably not well suited to numerical study.
However, as B\"{o}deker has shown, it has fairly simple behavior when
studied at the length scale $1/g^2T$ \cite{Bodeker}.  
In this regime the collision
integral dominates over the derivative term for the $W$ 
field\footnote{What follows is a gross oversimplification of
the argument, see \protect{\cite{moreBodeker,moreASY}}.}, since $C
\sim g^2 T \log(gT/\mu)$ with $\mu \sim g^2T$, while the derivative term 
is $v \cdot D \sim g^2T$; the collision term is therefore bigger by
$\sim \log(1/g)$.  Roughly speaking, one may solve for $W$ in terms of
$E$ in
Eq. (\ref{W_noH}) and plug it into Eq. (\ref{gauge_noH}), yielding a 
local expression of form $\sigma E = - \partial H/\partial A$.

At leading log order the Yang-Mills theory argument 
goes over directly to the case including a Higgs field, since it depends 
on manipulations of Eq. (\ref{W_noH}) only and this does not include the 
Higgs field.  The next to leading log calculation of \cite{AY2_long} also
still holds, provided there is no Higgs condensate of size $\phi_0 \gsim 
gT/\log(1/g)$.  We discuss this point at more length in an appendix,
where we show how the addition of the Higgs field does not modify the
calculation presented in \cite{AY2_long}.  Intuitively, the reason the
Higgs field does not affect $\sigma$ at next to leading log order is
as follows.  The conductivity depends on the efficiency of collisions.
The collisions are
mediated by gauge excitations with momenta $gT \gsim k \gsim g^2 T
\log(1/g)$.  At the low end of this range the Higgs fields modify the
gauge field thermodynamics by $O(1/\log(1/g))$, see Section
\ref{thermo}, and this part of the range gives a contribution down by
$1/\log(1/g)$ compared to the complete collision integral.  Hence the
influence of the Higgs field is suppressed by two powers of log.

Finally, we arrive at the effective theory
\bea
\label{eff_theory}
\sigma E_i^a(x) & = & - g^2T 
	\frac{\partial}{\partial A_i^a(x)} 
	\frac{H_{\rm eff}(A)}{T} + \xi_i^a(x) \, , \\
\label{Noise}
\langle \xi_i^a(x,t) \xi_j^b(y,t') \rangle & = & 
	2g^2T \sigma \delta_{ij} \delta_{ab} \delta^3(x-y) \delta(t-t')
	\, , \\ 
\label{value_of_sigma}
\sigma^{-1} & = & \frac{3}{m_D^2} \gamma \, , \qquad
	\gamma = \frac{N_{\rm c} g^2T}{4\pi} \left[ \ln \frac{m_D}{\gamma}
	+ 3.041 \right] \, .
\eea
Here $H_{\rm eff}$ is as in Eq. (\ref{def_Heff}) and the expression for
$\sigma^{-1}$ is from \cite{AY2_short} with a particularly nice choice for the 
renormalization scale $\mu$.  The sole modification the inclusion
of the Higgs fields has made is in the form of $H_{\rm eff}$.  Again we
emphasize that the the derivation has assumed that the Higgs is light
and that there is not a large Higgs condensate.  If there is a large
Higgs condensate $\phi_0 \gsim gT/\log(1/g)$ 
then the value of $\sigma$ changes, and if 
$\phi_0 \sim T$ then the form of the effective theory changes as well.
We will not discuss the latter case here.
Also note that our effective theory is only valid for studying gauge
field correlators; it cannot tell us much about unequal time Higgs
field correlators.

\section{Numerics}
\label{numerics}

B\"{o}deker's effective theory is Eq. (\ref{eff_theory}), but with
$H_{\rm eff}$ replaced with $H_A$.  It makes an excellent starting point
for numerical investigation of $\Gamma$ in pure Yang-Mills theory for
two reasons:  
\begin{enumerate}
\item
it is local, and
\item
it is UV finite.
\end{enumerate}
Neither is true for the effective theory we have derived.  This
potentially makes its study much more problematic than B\"{o}deker's
effective theory.

In practice the second problem is not a substantial one.  The theory is
not UV finite because $H_{\rm eff}$ involves the 3 dimensional path
integral for a scalar field, which contains linear and logarithmic mass
divergences arising from one and two loop graphs.  However, the path
integral is still super-renormalizable, and the UV infinities are purely 
local and can be absorbed with a mass counterterm.  Their value is
known \cite{Laine,LaineRajantie}, and as we will discuss below, we could 
actually proceed even if they were not.

The real problem with implementing the effective theory is its
nonlocality, which comes about because of the path integral in the
expression for $H_{\rm eff}$, see Eq. (\ref{def_Heff}).  The solution is 
to think about how we would carry out such a path integral numerically.
To evaluate Eq. (\ref{def_Heff}) numerically, 
we would perform the path integral by
Monte-Carlo, for instance by a Langevin equation,
\beqa
\frac{d\Phi}{d\tau_\phi}(x) & = & - \frac{\partial H_\Phi}
	{\partial \Phi^\dagger(x)} + \xi(x) \, , \\
\langle \xi(x,\tau_\phi) \xi^\dagger(x',\tau_\phi') \rangle & = & 
	2 T \; {\bf 1} \delta(x-x') \delta(\tau_\phi - \tau_\phi') \, .
\eeqa
Here the ${\bf 1}$ reminds us to make the noise diagonal in the
components of the Higgs field.
This Langevin equation must be evolved, and the result averaged, at
every time step in the Langevin dynamics of the gauge fields.

This suggests that Eq. (\ref{eff_theory}) can be replaced with the
$\eta \rightarrow \infty$ limit of the following system of equations:
\beqa
\label{really_use}
\sigma E_i^a(x) & = & - g^2 \frac{\partial}{\partial A_i^a(x)}
	( H_A(A) + H_\Phi(A,\Phi) ) + \xi_i^a(x,t) \, , \\
\sigma D_t \Phi(x) & = & - \eta \frac{\partial}{\partial
	\Phi^\dagger(x)} H_\Phi(A,\Phi) + \xi_\Phi(x,t) \, , \nonumber \\
\langle \xi_\Phi(x,t) \xi^\dagger_\Phi(x',t') \rangle 
	& = & 2 \eta \sigma T 
	\; {\bf 1} \delta(x-x') \delta(t-t') \, . \nonumber
\eeqa
This is Langevin evolution but with a different rate for the evolution
of the two fields, $A$ and $\Phi$.  In the limit that the Higgs field
evolution is made infinitely fast, which is $\eta \rightarrow \infty$, 
the Higgs field evolution will perform the path integral in a much
shorter time scale than the gauge fields evolve in, and
we recover the desired equations of motion, Eq. (\ref{eff_theory}).

Eq. (\ref{really_use}) at a finite value of $\eta$ does make a good
starting point for numerical work.  The large $\eta$ limit must then
be taken numerically.  It is also convenient to rescale time to a
Langevin time, $\tau \equiv \sigma t$.  Note that the
dimensions of $\tau$ are those of a length squared.

Our nonperturbative regularization for Eq. (\ref{really_use}) will be
the lattice.  Since all of the numerical tools we use exist in the
literature, we will only present the relevant references here rather
than give complete details.  The lattice discretization is standard, see
\cite{KLRSresults}.  Our topological lattice definition of $\NCS$
is the same as in \cite{MooreRummukainen}.  The relation between the
couplings of the lattice and continuum systems has been worked out in
\cite{Oapaper}, and we use the expressions there.  These relations match 
all thermodynamic quantities at $O(a)$, leaving $O(a^2)$ errors, with two 
exceptions.  We do not know the full $O(a)$ match for an additive piece
of the $\phi^2$ operator insertion or for $m_H^2(T)$.  If we were
interested in determining $\Gamma$ at a particular, fixed value of
$m_H^2(T)$ then this could pose a problem.  However, what we want is
$\Gamma$ when $m_H^2(T)$ is a fixed distance from the value which gives
phase equilibrium, $\meq^2$.  To determine this we do not need the
absolute normalization of $m_H^2(T)$ to $O(a)$, but we will 
need to find $\meq^2$ numerically.

Now we briefly discuss algorithm.  To determine $\meq^2$ we use the same 
algorithm and multicanonical techniques as \cite{KLRSresults}.  To
perform Eq. (\ref{really_use}), we may use whatever Higgs update we
choose, provided that it is stochastic and that we take the $\eta
\rightarrow \infty$ limit.  We choose a mixture of heat bath Higgs
updates and the x-y over-relaxation algorithm of \cite{KLRSresults}.
For the gauge field update we should use Langevin dynamics or any other
strictly dissipative dynamics.  We choose heat bath.  For either
algorithm, the effective infrared dynamics should be of the Langevin
form, and the difference between the algorithms will be a radiative
rescaling of the (Langevin) time scale and high dimension corrections,
which first appear at $O(a^2)$\footnote{The question of what
modifications may occur in the effective IR behavior has been addressed
at some length by Arnold and Yaffe, see particularly Appendix A of
\protect{\cite{AY2_first}}}.  The radiative rescaling of the time scale
is discussed at length in Appendix A of \cite{bodek_paper}.  
That paper presents an
analytic calculation of the lattice to continuum match for the Langevin
algorithm, and the Langevin to heat-bath rescaling is found numerically.
Within error it is equal to the coefficient $Z_g^{-1}$ which appears in
the $O(a)$ thermodynamical match detailed in \cite{Oapaper}.  We will
assume that this numerically determined relation holds analytically.

\section{Results}
\label{result_sec}

With a numerical implementation of the effective theory in hand we will
address two questions.  First, how large is $\Gamma$ in the symmetric
phase, at $\meq^2$, for parameters where the phase transition is strong
enough to preserve baryon number after its completion?  Second, how does 
$\Gamma$ vary as we go through the analytic crossover present for large
vacuum Higgs masses?

\begin{figure}[tbp]
\centerline{ \mbox{\epsfxsize=3in\epsfbox{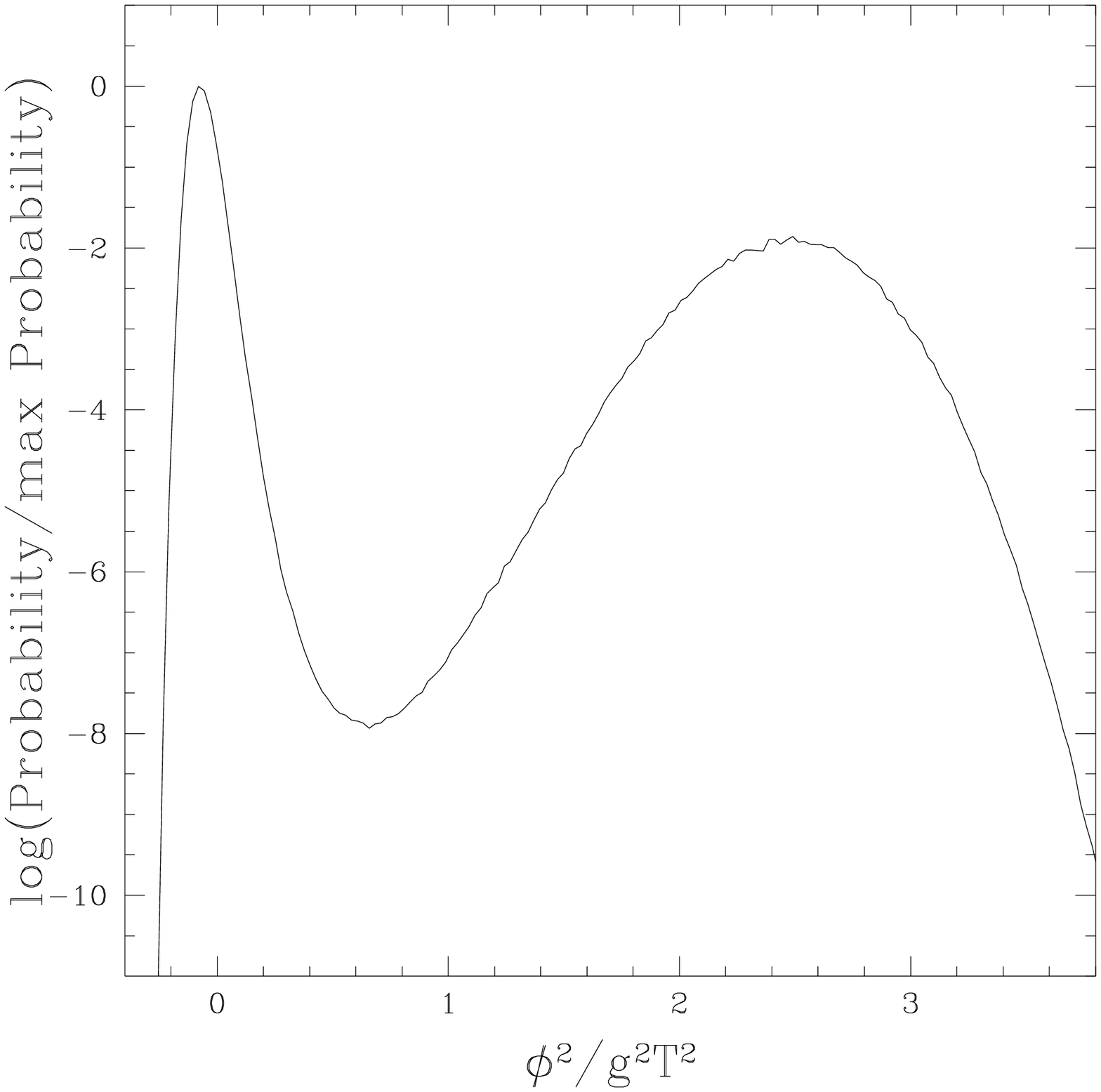}} \hspace{0.2in}
	     \mbox{\epsfxsize=3in\epsfbox{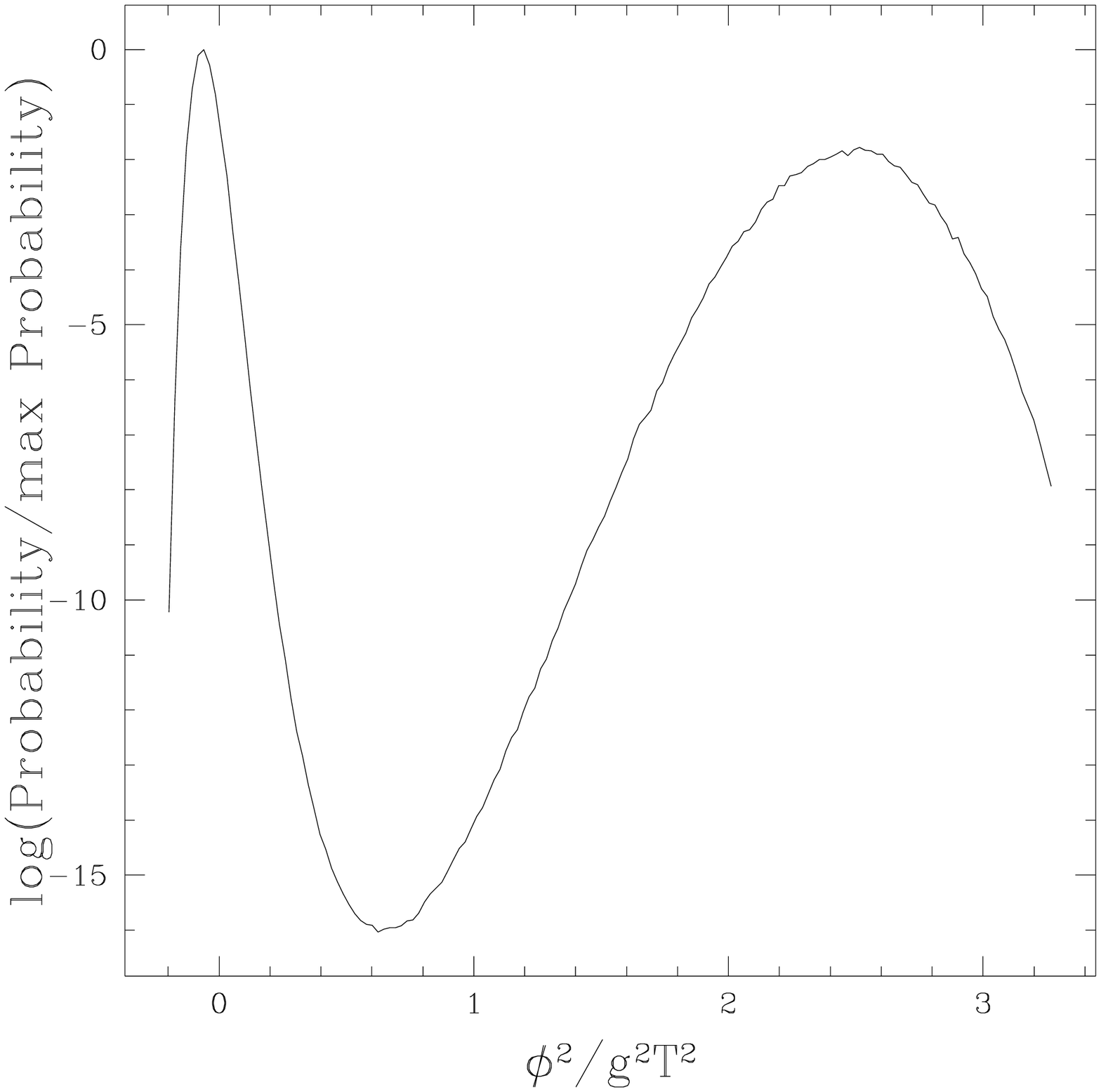}} }
\caption{\label{prob_dist} Log of probability distribution for $\phi^2
\equiv 2 \Phi^\dagger \Phi$ ($\overline{\rm MS}$ renormalization point
$\overline{\mu}=g^2T$) for cubic boxes of size $24^3$ (left) and $32^3$
(right), at lattice spacing $4/9g^2T$ ($\beta=9$).  Inverting the $y$
axis roughly gives the free energy dependence on $\phi^2$.  
The left peak is the 
symmetric phase and the right peak is the broken phase; the dip between
them is a free energy barrier separating the phases, which grows as the
volume is increased (it is not free energy per volume).  The asymmetric
shape is typical for this order parameter.}
\end{figure}

To address the first question we consider Yang-Mills Higgs theory at
$\lambda / g^2 = 0.036$, which is roughly the value at which the phase
transition is barely strong enough to preserve baryon number
cosmologically after its completion.\footnote{The value we use is
slightly smaller than that found in \protect{\cite{broken_nonpert}}
because here we do not include the U(1) field.}  We determine $\meq^2$
at three lattice spacings, $a=4/7g^2T$, $a=4/9g^2T$, and $a=4/12g^2T$
(which in the notation which is sometimes customary \cite{AmbKras} are
$\beta = 7$, 9, and 12), by multicanonical Monte-Carlo.  The probability 
distributions for $\phi^2$ at $\meq^2$ for 
two lattice volumes and $a = 4 / 9g^2T$ are
shown in Fig. \ref{prob_dist}.

First we consider the $\eta \rightarrow \infty$ limit at fixed lattice 
spacing $a=4/9g^2T$ in a cubic volume $L^3$ 
with periodic boundary conditions, for
$L = 14.2/g^2T$ (32 sites on a side).  This
volume is abundantly large enough to see the large volume value of
$\Gamma$ \cite{MooreRummukainen}, and is large enough that strong
metastability prevents tunneling to the broken phase at $\meq^2$, where
we work.  Table \ref{table1} presents $\kappa'$, defined in
Eq. (\ref{parametric_form}), for various values of $\eta$, lattice
spacing, and $m_H^2$.  The $\eta$ dependence is weak and statistically
compatible with zero.

\begin{table} 
\centerline{ \begin{tabular}{|c|c|c|c|c|}\hline
$\; a \times g^2T \;$ & $\quad \eta \quad$ & $\; \; m_H^2-\meq^2
\; \; $ & ~volume$\times$Langevin time~ 
& $\quad \qquad \kappa' \quad \qquad $ \\ \hline
4/9  &   2.5 &  0       &  $32^3\times 56400$  &  $8.45 \pm .24$ \\ \hline
4/9  &   5   &  0       &  $32^3\times 56400$  &  $8.20 \pm .24$ \\ \hline
4/9  &   10  &  0       &  $32^3\times 56400$  &  $8.08 \pm .23$ \\ \hline
4/9  &   5   &  .0093   &  $32^3\times 34300$  &  $8.63 \pm .32$ \\ \hline
4/9  &   5   &  -.0093  &  $32^3\times 30900$  &  $8.50 \pm .33$ \\ \hline
4/7  &   10  &  0       &  $24^3\times 57400$  &  $8.13 \pm .37$ \\ \hline
4/12 &   5   &  0       &  $40^3\times 98600$  &  $7.83 \pm .27$ \\ \hline
4/9  & Yang  &  Mills   &  $24^3\times 489400$ &  $9.90 \pm .13$\\ \hline
4/10 & Yang  &  Mills   &  $40^3\times 84900$ &  $10.00 \pm .23$\\ \hline
\end{tabular}}
\vspace{0.15in}
\caption{Chern-Simons number diffusion $\Gamma$ in the symmetric phase
near the equilibrium temperature with $\lambda/g^2 = .036$, for
different lattice spacings, values of $\eta$, and Higgs masses.  The
4-volume is expressed in lattice units; to convert to physical units
multiply lengths by $a$ and Langevin time by $\sigma a^2$.  Except
for the pure Yang-Mills data, all values of $\kappa'$ are statistically
compatible; no lattice or $\eta$ dependence is statistically
significant. \label{table1}}
\vspace{0.1in}
\end{table}

\begin{figure}
\centerline{\epsfxsize=3in\epsfbox{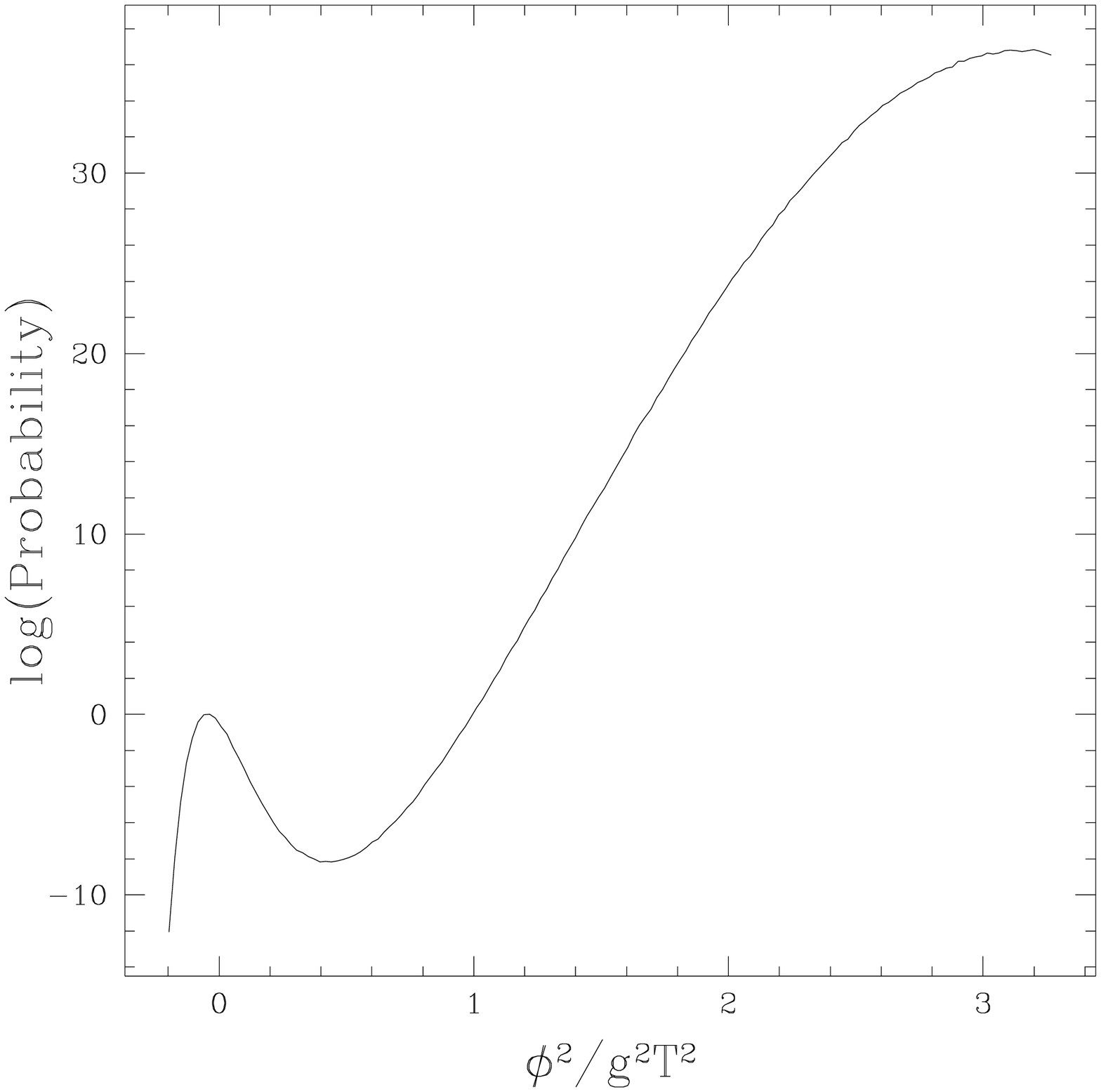} \hspace{0.2in}
\epsfxsize=3in\epsfbox{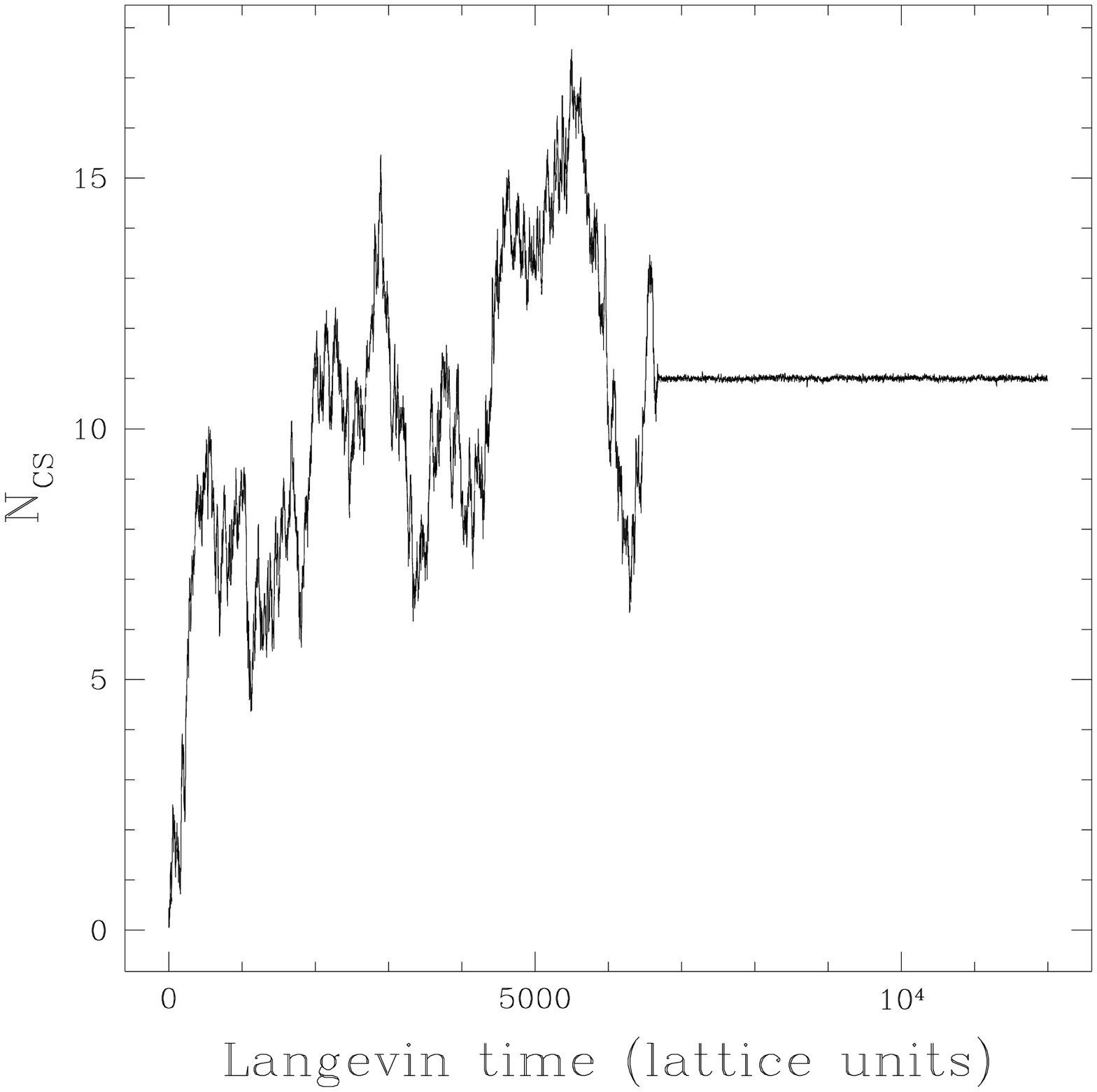}}
\caption{\label{supercool}
Left:  probability distribution on supercooling, $m_H^2 = \meq^2 - .0093 
g^4T^2$.  Right:  Langevin time history of $\NCS$ starting in the
symmetric phase, under such supercooling.  After a tunneling event to
the broken phase, $\NCS$ stopped diffusing.}
\end{figure}

Table \ref{table1} also shows that, as expected, there is almost no
lattice spacing dependence in $\Gamma$.  The $O(a)$ match for the
thermodynamic quantities and time scales is essential here, see
\cite{bodek_paper}.  Finally, since in the usual electroweak
baryogenesis scenario the symmetric phase undergoes supercooling, we
should study $\Gamma$ in the supercooled symmetric phase.  Table
\ref{table1} shows that the maximum supercooling compatible with strong
metastability is still not enough to significantly change $\Gamma$.
We cannot increase the supercooling beyond what was used because the
lattice system will nucleate to the broken phase.  Indeed, one run used
for the table ended with a nucleation to the broken phase, as is seen
clearly from the time history for $\NCS$ in that run, shown in
Fig. \ref{supercool}.  Naturally, the broken phase part of the evolution 
was not used in the analysis.
Combining all the figures in the table, since all 
are statistically compatible and no trend (lattice spacing,
supercooling, or $\eta$) is statistically significant, we get
\begin{equation}
\Gamma_{\rm symm} \simeq \big[ 8.24 \pm 0.10 \big] 
	\left( \frac{g^2T^2}{m_D^2}
	\right) \left( \log \frac{m_D}{g^2T} + C \right) 
	\alpha^5 T^4 \, ,
\end{equation}
for $\lambda / g^2 = 0.036$.  We have computed the pure Yang-Mills
theory value at the same lattice spacing with comparable precision to
facilitate comparison; the symmetric phase value is $0.832 \pm .015$ of
the Yang-Mills theory value.  (The Yang-Mills theory value found here 
is smaller than that quoted in \cite{bodek_paper}, where we found
$\kappa' = 10.8 \pm 0.7$.  This is a statistical fluctuation in the data
there.  The table includes a higher statistics redetermination of
$\kappa'$ at the same parameters used in that paper, with a result in
agreement with the $\beta=9$ value found here, 
and statistically compatible at $1\sigma$ with the result determined in
\cite{bodek_paper}.)

\begin{figure}[tbp]
\centerline{ \mbox{\epsfxsize=3in\epsfbox{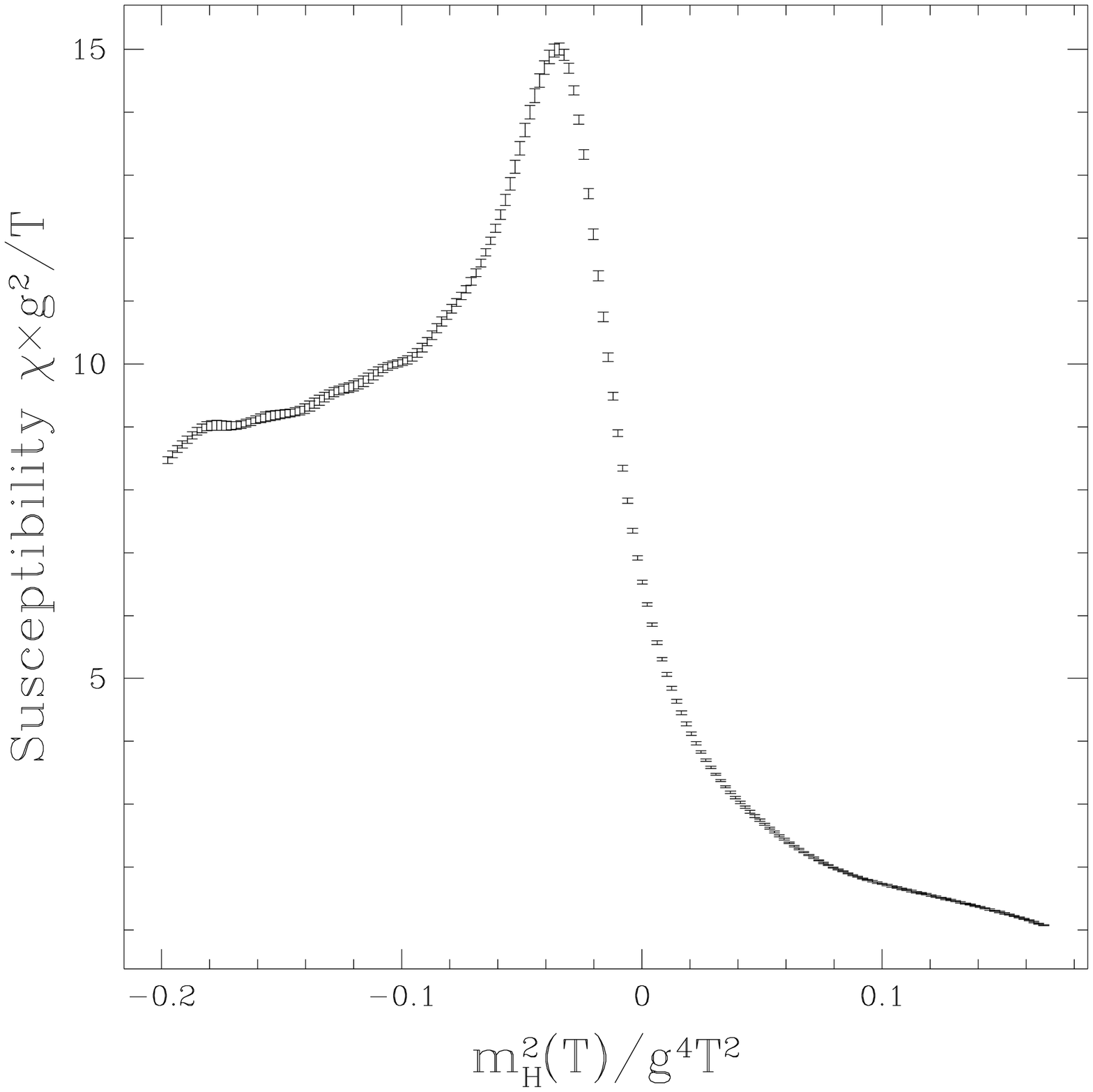}} \hspace{0.2in}
	     \mbox{\epsfxsize=3in\epsfbox{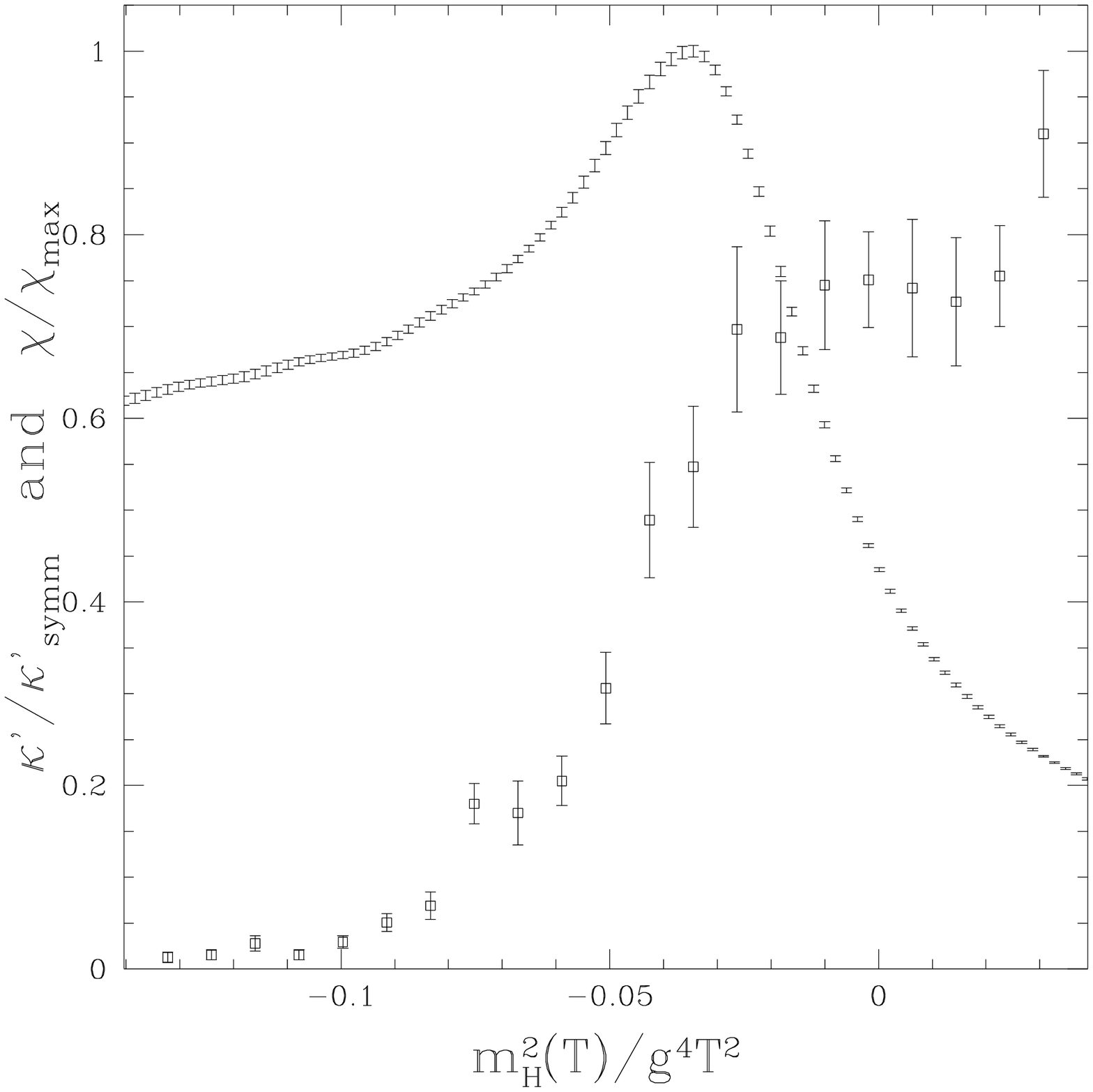}} }
\caption{\label{cross_fig} Left:  
$\phi^2$ susceptibility; Right:  $\phi^2$ susceptibility (peak with
small errors) and sphaleron rate (larger errors, square plotting
symbols) scaled to their maxima,
when there is a smooth crossover.  $m_H^2(T)$ plotted is the 3D theory
value, $\overline{\rm MS}$ renormalized with $\overline{\mu}=g^2T$.}
\end{figure}

Of course the above is academic because $\lambda / g^2 = 0.036$ converts 
to a vacuum Higgs mass $m_H = 44$GeV at tree level.  Including the large 
radiative top Yukawa corrections, this $\lambda / g^2$ does not
correspond to any physical Higgs mass \cite{KLRS}.  In the standard
model, all experimentally allowed Higgs masses fail to provide an
electroweak phase transition at all.  Rather, there is an analytic crossover
\cite{KLRS_cross}.
Although baryogenesis is probably impossible in that setting, it would
still be interesting to see how $\Gamma$ varies as we go through the
crossover.  Does it turn on suddenly or gradually, and does its turn-on
point coincide with the peak in the susceptibility for $\phi^2$?  We
answer this question in Fig. \ref{cross_fig}, which shows how the
$\phi^2$ susceptibility 
\begin{equation}
\chi_{\phi^2} \equiv \frac{1}{V} \left[ \left\langle \left( \int \phi^2 
	\right)^2 \right\rangle - \left\langle \int \phi^2 
	\right\rangle^2 \right] \, ,
\end{equation}
and the sphaleron rate $\Gamma$ vary with $m_H^2(T)$.  The data are for
$\lambda / g^2=5/16$, corresponding at tree level to a physical vacuum
$m_H = 130$GeV, and were taken for a $32^3$ box with lattice spacing $a
= 1/2g^2T$.  The sphaleron rate data were all taken using $\eta=20$.
We see that the
switch-on of $\Gamma$, though smooth, is fairly rapid and occurs at
slightly lower $m_H^2(T)$ than
the peak susceptibility; that is, it is when conditions are a little
more ``broken phase-like'' than when the susceptibility peaks.  The
sphaleron rate $\Gamma$ proves a rather good order parameter to
distinguish Higgs-like and symmetric-like phases.  Although, like any
order parameter must, it shows smooth behavior, the range where it is
far from both its broken phase value $\kappa' \simeq 0$ and its
symmetric phase value $\kappa'>5$ is quite narrow, roughly as narrow as
the peak in the $\phi^2$ susceptibility.  However, we do not view the
sphaleron rate as
competitive with the susceptibility as a probe of where the crossover
occurs.  The main reason is that it is possible to perform a
multicanonical reweighting which allows the susceptibility to be scanned 
in a wide range of $m_H^2$ (the data in the figure come from a single
numerical run), and the statistics for the susceptibility improve more
quickly.  For instance, the $\NCS$ diffusion data in
Fig. \ref{cross_fig} took more CPU time than the susceptibility data,
but are substantially ``dirtier''.  (The data sets are plotted with
their error bars.  The $\NCS$ diffusion errors are each independent, 
but there is very large cross-correlation in errors of neighboring
points for the susceptibility because they were all computed from one
data set.)

\section{Conclusion}

Assuming that classical 3-D scalar $\phi^4$ theory is ergodic, 
the addition of a light Higgs degree of freedom replaces B\"{o}deker's
effective theory for the evolution of infrared gauge fields with a
slightly more complicated equation, Eq. (\ref{eff_theory}).  The sole
change is in the thermodynamic potential for the gauge fields.
Including the Higgs field makes this thermodynamic potential nonlocal.
However, the effective theory 
is still a useful starting point for numerical work because
we can use the limit of a sequence of local effective theories, namely,
Langevin (or heat bath) evolution for the gauge fields and for the Higgs 
fields, but with much faster Langevin evolution for the Higgs degrees of 
freedom.

When the electroweak phase transition is strong, the sphaleron rate in
the symmetric phase is reduced by around $20\%$ from its Yang-Mills
theory value, roughly in accord with an estimate based on the 
thermodynamics.  (The thermodynamic estimate for small $\lambda / g^2$
is $(\delta \Gamma / \Gamma) \simeq 1.2 \sqrt{\lambda/g^2}$, see
Eq. (\ref{dgamma}).)   
When there is no electroweak phase transition,
but an analytic crossover, the sphaleron rate changes rather quickly
from its symmetric phase value to nearly zero, roughly at the same value 
of $m_H^2(T)$ where the $\phi^2$ susceptibility peaks.  The crossover
region is of about the same width as the peak in the $\phi^2$
susceptibility and is displaced to slightly lower $m_H^2(T)$.

In this paper we have only studied the standard model, either for
parameters which are ruled out experimentally or for which the
electroweak phase transition cannot provide for baryogenesis.  However
it is fairly simple to see how to extend the work to more viable models
like the MSSM (minimal supersymmetric standard model).  
In that case, the SU(2) and SU(3) gauge fields would
each evolve under Langevin dynamics, with a thermodynamic potential
arising from integrating over the Higgs and scalar top fields.  However, 
from our results with a strong phase transition, it should be clear that,
in all cases where the phase transition is strongly first order,
$\Gamma$ in the symmetric phase will be very close to its Yang-Mills
theory value.  This is because we have found that the suppression of
$\Gamma$ corresponds well to what we expect thermodynamically; and when
the electroweak phase transition is strong in the MSSM, the symmetric
phase Higgs mass is larger than in the standard model case.  Therefore
the SU(2) thermodynamics in the symmetric phase is closer to Yang-Mills 
theory in the MSSM than in the standard model, see
Eq. (\ref{heavy_Higgs}).  Hence it is almost
certainly true that, in the MSSM and when the phase transition is
strong, the symmetric phase sphaleron rate is lower than but 
within $20\%$ of the pure Yang-Mills theory value.  In practice this
means we can continue to quote the Yang-Mills theory result for all
symmetric phase cases of physical interest, with modest error.

\section*{Acknowledgments}

I am grateful to Dam Son for a useful discussion in which he conjectured 
that the Higgs fields would enter only through $H_{\rm eff}$.  I also
acknowledge useful conversations with Larry Yaffe and Dietrich
B\"{o}deker.  This work 
was partially supported by the DOE under contract DE-FGO3-96-ER40956.

\appendix

\section{Higgs fields and Hard Thermal Loops}
\label{App_HTL}

In this appendix we show that the only hard thermal loop needed for the 
bosonic effective theory, which has soft external scalar field lines,
is the scalar mass correction.

We need only consider HTL's with bosonic external lines, since the
appropriate IR effective theory we seek is bosonic.  Hence we need in
general to consider all diagrams with $(n_s>0,{\rm even})$ scalar and
$n_g$ gauge boson external lines.  We will use repeatedly the power
counting rules derived in \cite{BraatenPisarski}, which we repeat here
for the reader's convenience.  A hard thermal loop is always an $O(T^2)$ 
contribution from the $K \sim T$ momentum region of a one loop
integral with soft ($gT$) external lines.  To determine the largest
power of $T$ possible from a diagram, follow these rules:
\begin{enumerate}
\item the loop integration contributes $T^3$;
\item the first propagator times the Matsubara sum contributes $1/T$;
\item every additional propagator contributes $1/(PT)$;
\item powers of $K^\mu$ in the numerator (from 3 point gauge vertices or 
fermionic propagators) each contribute $T$;
\item when there are two or more propagators, and all propagators are
either bosonic or fermionic, there is an extra $P/T$ suppression.
\end{enumerate}
If the result has a weaker power of $T$ than $T^2$ the diagram does not
contribute an HTL; if it is $T^2$ the diagram will unless there is some
cancellation.  

\begin{figure}[t]
\centerline{\epsfxsize=6in\epsfbox{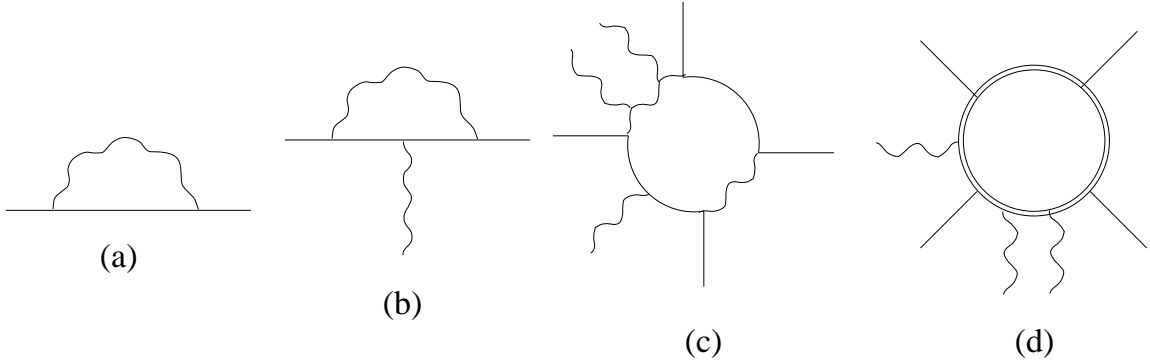}}
\vspace{0.1in}
\caption{Diagrams considered in Appendix \protect{\ref{App_HTL}}.  Solid 
lines are scalars, wiggly lines are gauge bosons, double lines are
fermions.  Diagrams (a) and (b) are a self-energy and a vertex
correction considered in the text.  Diagrams (c) and (d) are generic
diagrams with external gauge and scalar lines and either a boson
(alternating Higgs and gauge, diagram (c)) or a fermion (diagram (d)) in 
the loop.  As discussed, (b), and all diagrams of type (c) and (d), do
not give rise to hard thermal loops. \label{fig_HTL}}
\end{figure}

Since we are only concerned with HTL's with bosonic external legs, then
either all the propagators in the loop will be bosons or all will be
fermions.  Any diagram with more than 1 propagator will get a $P/T$ from 
rule 5.~and this will exclude diagrams with 4 point vertices (4 gluon, 2 
gluon and 2 scalar, or 4 scalar) except for tadpoles, just as is the
case in Yang-Mills theory.  This leaves all
diagrams of the general form of diagrams (c) and (d) in Figure
\ref{fig_HTL}.  

The scalar self-energy diagrams are considered in \cite{Thoma} (fermion
loop) and \cite{Kraemmer} (boson loops).  The result is that the
self-energy is a momentum independent mass correction.  In particular it 
is useful to reproduce the argument for the gauge HTL from diagram (a)
of Figure \ref{fig_HTL}.  In Feynman gauge this diagram contributes
(following the notation of that paper, where capital letters are
4-vectors and lower case are spatial or temporal (0 subscript)
components) 
\begin{eqnarray}
&& - \frac{N^2-1}{2N}g^2 T \sum_{k_0} \int \frac{d^3k}{(2\pi)^3} 
	\frac{(2P+K)^\mu (2P+K)^\nu g_{\mu \nu}}{K^2 (K+P)^2} \nonumber \\
& = & -\frac{N^2-1}{2N}g^2T  T \sum_{k_0} \int \frac{d^3k}{(2\pi)^3} 
	\left[\frac{2(K+P)^2 - K^2 + 2P^2}{K^2(K+P)^2} \right]\, .
\end{eqnarray}
For the first term in the square brackets, the $(K+P)^2$ in the
numerator cancels a propagator in the denominator, leaving one
propagator and no powers of $K$ in the numerator.  This gives an HTL
effect only because there is a single propagator left, so 5.~above is not
invoked.  The result is a momentum independent tadpole.  
The second term is similar.  The
last term has two propagators and no power of $K$ in the numerator; it
is $O(T^0)$ and does not give an HTL.

Similarly, diagram (b) gives, in Feynman gauge and labeling the incoming 
scalar momentum $P_1$, the incoming gauge momentum $P_2$, 
and the outgoing scalar momentum $P_3 = P_1 + P_2$, 
\begin{equation}
\propto g^3 T \sum_{k_0} \int \frac{d^3k}{(2\pi)^3} 
	\frac{(K+2P_1)\cdot(K+2P_3) (2K+P_1+P_3)^\mu}
	{ K^2 (K+P_1)^2 (K+P_3)^2 } \, .
\end{equation}
It is convenient to rewrite
\begin{equation}
(K+2P_1) \cdot (K+2P_3) = (K+P_1)^2 + (K+P_3)^2 - K^2 
	+(4P_1 \cdot P_3 - P_1^2 - P_3^2) \, .
\end{equation}
Each $K^2$ type term will cancel a propagator, leaving two propagators
and one $K^\mu$ in the numerator.  By the power counting rules this
contributes at most $T^3 (1/T)(1/PT) T (P/T) \propto T^1 $ and does not 
give a hard thermal loop.  The $P^2$ term is $O(T^0)$, so it does
not either.

This example illustrates why diagrams of form (c) and (d), with $(n_s
>0, {\rm even})$ scalars and $n_g$ gauge bosons, $(n_s+n_g)>2$, do
not give hard thermal loops.  Such a diagram has $(n_s + n_g)>2$
propagators.  It also has $n_s + n_g$ factors of $K^\mu$ in the
numerator, either all from gauge vertices (bosonic loops) or all from
fermionic propagators (fermionic loops).  However, there are only
$n_g$ external Lorentz indices.  In Feynman gauge, the diagram will be a sum of
terms, with at least $n_s$ of the $K^\mu$ contracted against each other
in each term.  Hence, each term from each diagram contains in its
numerator, in Feynman gauge, a term of form $(K+P_i) \cdot (K+P_j)$,
with $P_i,P_j$ some linear combinations of external momenta and $K$ the
loop momentum.  This can always be written as $\sum ({\rm sign})
(K+P_l)^2 + O(P^2)$, with each $(K+P_l)$ a momentum on some propagator.
Consider one of the $(K+P_l)^2$ terms.  Cancelling the $(K+P_l)^2$
against the appropriate propagator leaves $(n_s+n_g-2)$ $K^\mu$ factors
in the numerator but $(n_s+n_g-1)>1$ propagators.  The power counting then 
gives at most $T^3 (1/T) (1/PT)^{n_s+n_g-2} T^{n_s+n_g-2} (P/T)
\propto T^1$, too small to contribute an HTL, even without further
cancellations.  Meanwhile, the $O(P^2)$ term has $(n_s+n_g)$ propagators 
but $(n_s+n_g-2)$ $K^\mu$ factors; it is at most 
$P^2 T^3 (1/T)(1/PT)^{n_s+n_g-1} T^{n_s+n_g-2} (P/T) \propto T^0$ and
also gives no HTL.  Though we worked in Feynman gauge, we expect the
hard thermal loops to be gauge invariant; if they vanish except a
momentum independent scalar mass in Feynman gauge, they should in any
gauge.

Hence the only bosonic HTL with scalar external lines is the gauge field 
mass insertion.  Note that similar reasoning to what we give above also
rules out HTL's with both scalar and fermionic external lines; but this
is unimportant for the purposes of this paper.

\section{Higgs fields and next to leading log conductivity}

In this appendix we briefly explain how the calculation presented in
\cite{AY2_long} is modified by the presence of Higgs fields, and why the 
modification does not change the conductivity of the final effective
theory at next to leading log order.  The purpose of the appendix is to
outline the argument, at times we will be sloppy with notations and with 
terms which are not relevant and have been discussed at much more length 
in \cite{AY2_first,AY2_short,AY2_long}.

The relevant part of the calculation in \cite{AY2_long} is the match
between their ``theory 2'' and ``theory 3'', only now ``theory 2'' is
given by Eqs. (\ref{gauge_noH}-\ref{W_noH}), which for convenience we
repeat here:
\bea
\label{first_eq}
\mD \int \frac{d\Omega}{4\pi} v_iW^a(x,v) & = & - g^2T 
	\frac{\partial}{\partial A_i^a(x)} 
	\frac{H_{\rm eff}(A)}{T} \, , \\
\frac{H_{\rm eff}(A)}{T} & = & - \log \int {\cal D}\Phi \exp 
	\left( - \frac{H_A + H_{\Phi}}{T} \right) \, , \\
v_i (D_i W)^a(x,v) & = & \mD v_i E_i^a(E) 
	- \int \frac{d\Omega_{v'}}{4\pi} C_{vv'} W^a(x,v') 
	+ \zeta^a(x,v) \, .
\label{W_eq}
\eea

It is possible, at least
formally, to invert Eq. (\ref{W_eq}) to solve for $W$ in terms of $E$.
Eq. (\ref{first_eq}) then becomes
\begin{equation}
(\sigma_{ij}(D) E_j)(x) =  - g^2T 
	\frac{\partial}{\partial A_i(x)} 
	\frac{H_{\rm eff}(A)}{T} + \zeta_i ' \, ,
\end{equation}
where $\sigma(D)$ is the nonlocal operator resulting from inverting
Eq. (\ref{W_eq}); it is discussed in \cite{AY2_long}.  $\zeta'$ is a
noise with correlator $\langle \zeta' \zeta' \rangle = 2T \sigma(D)$.
(Starting here we will aggressively suppress indices where we feel the
meaning is clear.)

This Langevin equation only has a path integral representation if we are 
willing to accept $(\delta H_{\rm eff}/\delta A)^2$
inside the action of a path integral.  It is not clear how to form a
perturbation theory for such a path integral.  However, if we rewrite
$H_{\rm eff}$ as we did in Section \ref{numerics}, then it will become
possible.  The effective theory of interest is the $\eta \rightarrow
\infty$ limit of
\bea
\sigma_{ij}(D) E_j & = & -g^2T \left( \frac{\partial H_A}{\partial A_i}
	+ \frac{\partial H_\Phi}{\partial A_i} \right) \, , \\
D_t \Phi & = & - \eta \frac{\partial H_\Phi}{\partial \Phi^\dagger}
	+ \xi \, , \\
\langle \xi(x,t) \xi^\dagger (x',t') \rangle & = & 2\eta T {\bf 1}
\delta(x-x') \delta(t-t') \, .
\eea
To write a path integral expression for this, we follow the standard
trick of writing the Langevin equation and the average over the noise
distribution as a path integral,
\begin{eqnarray}
\int {\cal D} \zeta' {\cal D} \xi {\cal D} A && {\cal D} \Phi
	 \; \exp \left(- \int \zeta' 
	\frac{1}{4g^2\sigma(D) T} \zeta' \right) \; 
	\exp \left(-\int \xi \frac{1}{4\eta T} \xi \right) \nonumber \\
&& \times \delta \left(\zeta' + \sigma(D) E 
	+ g^2T \left[\delta (H_A+H_\Phi)/\delta A \right] \right)
	\delta \left( \xi + D_t \Phi + \eta \left[ \delta H_\Phi / 
	\delta \Phi^\dagger \right] \right) \, ,
\end{eqnarray}
times Jacobians for each field, which are not important, 
see \cite{AY2_first,AY2_long}.
Here the path integral performs the average over the noise, and the
delta functions enforce the Langevin equations.  Enforcing the delta
function does the integrals over each noise, giving a path integral
\begin{eqnarray}
&& \int {\cal D}A {\cal D}\Phi \exp (-L) \, , \nonumber \\
L & = & \left(\sigma(D) E 
	+ g^2T \left[\frac{\delta (H_A+H_\Phi)}{\delta A} \right] 
	\right) \frac{1}{4g^2 \sigma(D) T} 
	\left(\sigma(D) E 
	+ g^2T \left[\frac{\delta (H_A+H_\Phi)}{\delta A} \right]
	\right) \nonumber \\
&& + \left(  D_t \Phi + \eta \left[ \frac{\delta H_\Phi}
	{\delta \Phi^\dagger} \right] \right)^\dagger
	\frac{1}{4 \eta T}
	\left(  D_t \Phi + \eta \left[ \frac{\delta H_\Phi}
	{\delta \Phi^\dagger} \right] \right) \, .
\end{eqnarray}
Here we have not written the contributions from the Jacobians or from an 
extra regulation dependent term arising because of the nonlocality of
$\sigma$, which ref.~\cite{AY2_long} calls $L_1[A]$.  
These terms, and the reasons they can be dropped, are
discussed at some length in \cite{AY2_first,AY2_long}.

Arnold and Yaffe have shown that $\sigma$ in
Eq. (\ref{eff_theory}) can be determined at next to leading log order by
computing the $\omega = 0$, $k \rightarrow 0$ limit of the one loop, 
Coulomb gauge $A_0A_0$ self-energy for the theory with $\sigma(D)$,
and finding what constant value of $\sigma$ is needed to get the same
one loop result in the final 
effective theory.  To account for Higgs contributions in this
calculation, we must find all one loop Higgs field self-energy
corrections to the $A_0$ field, and must determine whether their
contributions survive in the $\eta \rightarrow \infty$ limit.

First we have to find what vertices couple $\Phi$ to $A_0$.
The only terms in the Lagrangian which contain $A_0$ at all 
are those which contain $E$ or $D_t$.  They are
\begin{equation}
\label{terms_weneed}
\frac{1}{4} E \sigma(D) E + \frac{1}{4 \eta} (D_t \Phi)^\dagger 
	(D_t \Phi)
\end{equation}
plus terms odd in $D_t$, which are of form
\begin{equation}
\frac{1}{2} E_i \frac{\delta H_A}{\delta A_i} \, , \qquad
\frac{1}{4} \left( 2 E_i \frac{\delta H_\Phi}{\delta A_i} + 
	(D_t \Phi)^\dagger \frac{\delta H_\Phi}{\delta \Phi^\dagger}
	+ cc \right) \, .
\end{equation}
The first term here is proportional to $\partial_t H_A$.  The second is 
proportional to $\partial_t H_\Phi$.  Both are total spacetime
derivatives, which can be converted to boundary terms at asymptotically
early and late times and therefore neglected.  Hence we only need to
consider vertices which arise from the terms in
Eq. (\ref{terms_weneed}).  These allow two new diagrams not considered
in \cite{AY2_long}, which are the same as (a) and (b) in Figure
\ref{diagrams} if the wavy lines are now taken to be $A_0$ propagators.

To show that these diagrams vanish in the small $\eta$ limit, we need
only count powers of $\eta$ in the vertices and propagators.  
The $A_0 \Phi^2$ vertex carries a factor of $\omega / \eta$ and the
$A_0^2 \Phi^2$ vertex carries a factor of $1/\eta$.  The $\Phi$ propagator 
is 
\begin{equation}
\label{phi_prop}
\langle \Phi \Phi \rangle = \frac{1}{ \omega^2/\eta + \eta k^4 } \, .
\end{equation}
The contribution from diagram (a) then scales as
\begin{equation}
{\rm (a)} \propto \frac{1}{\eta}
	\int d\omega d^3k \; \frac{1}{\omega^2 / \eta + \eta k^4} 
	\sim \eta^{-5/4} \, ,
\end{equation}
while diagram (b) behaves as 
\begin{equation}
{\rm (b)} \propto \int d\omega d^3k \; \frac{\omega^2}{\eta^2} \: \frac{1}
	{ (\omega^2/\eta + \eta k^4)^2 }
	\sim \eta^{-5/4} \, ,
\end{equation}
and we see that both vanish when we take $\eta \rightarrow \infty$.
Therefore there are no new contributions at one loop from Higgs bosons,
and the next to leading order calculation of $\sigma$ is unaffected.
This will not be true in the case were there is a Higgs condensate of
magnitude $\phi_0 \sim gT/\log(1/g)$, because in that case 
the external Higgs field insertions on the $A$ field lines in the one
loop diagrams considered in \cite{AY2_long} will change the gauge field
propagators by order 1.  The Higgs field will also be important beyond
one loop, in the construction of an effective theory which can determine 
the $O(1/\log)$ term in Eq. (\ref{parametric_form}); 
but we will not investigate that problem here.

\newcommand{\hep}[1]{[{#1}]}

\end{document}